\documentclass[twocolumn]{aastex631}
\usepackage{amsmath}
\newcommand{\msun}{\mathrm{M_{\odot}}} 
\newcommand{\msunyr}{\msun\,{\rm yr}^{-1}} 

\newcommand{\mdot}{\dot{M}} 
\newcommand{\mesa}{{\tt\string MESA}}

\newcommand{\kB}{k_{\mathrm{B}}}
\newcommand{\NA}{N_{\mathrm{A}}}

\newcommand{\Jdotgr}{\dot{J}_{\mathrm{gr}}}

\newcommand{\Porb}{P_{\mathrm{orb}}}
\newcommand{\Pcrit}{P_{\mathrm{c}}}

\newcommand{\Etide}{\dot{E}_{\mathrm{tide}}}
\newcommand{\Eheat}{\dot{E}_{\mathrm{heat}}}

\newcommand{\Teff}{T_{\mathrm{eff}}}

\newcommand{\Rmin}{R_{\mathrm{min}}}
\newcommand{\RWD}{R_{\mathrm{WD}}}

\submitjournal{ApJL}

\shorttitle{Orbital decay in an accreting and eclipsing 13.7 minute orbital period binary with a luminous donor}
\shortauthors{Burdge et al.}

\begin{document}

\title{Orbital decay in an accreting and eclipsing 13.7 minute orbital period binary with a luminous donor}

\correspondingauthor{Kevin B. Burdge}
\email{kburdge@mit.edu}

\author[0000-0002-7226-836X]{Kevin B. Burdge}
\affiliation{Department of Physics, Massachusetts Institute of Technology, Cambridge, MA 02139, USA}
\affiliation{Kavli Institute for Astrophysics and Space Research, Massachusetts Institute of Technology, Cambridge, MA 02139, USA}

\author[0000-0002-6871-1752]{Kareem El-Badry}
\affiliation{Center for Astrophysics $|$ Harvard \& Smithsonian, 60 Garden Street, Cambridge, MA 02138, USA}
\affiliation{Harvard Society of Fellows, 78 Mount Auburn Street, Cambridge, MA 02138}
\affiliation{Max-Planck Institute for Astronomy, K\"onigstuhl 17, D-69117 Heidelberg, Germany}

\author[0000-0003-3182-5569]{Saul Rappaport}
\affiliation{Department of Physics, Massachusetts Institute of Technology, Cambridge, MA 02139, USA}
\affiliation{Kavli Institute for Astrophysics and Space Research, Massachusetts Institute of Technology, Cambridge, MA 02139, USA}

\author[0000-0001-9195-7390]{Tin Long Sunny Wong}
\affiliation{Department of Physics, University of California, Santa Barbara, CA 93106, USA}

\author[0000-0002-4791-6724]{Evan B. Bauer}
\affiliation{Center for Astrophysics $|$ Harvard \& Smithsonian, 60 Garden Street, Cambridge, MA 02138, USA}

\author[0000-0001-8038-6836]{Lars Bildsten}
\affiliation{Department of Physics, University of California, Santa Barbara, CA 93106, USA}
\affiliation{Kavli Institute for Theoretical Physics, University of California, Santa Barbara, CA 93106, USA}

\author[0000-0002-4770-5388]{Ilaria Caiazzo}
\affiliation{Division of Physics, Mathematics and Astronomy, California Institute of Technology, Pasadena, CA 91125, USA}

\author[0000-0001-8804-8946]{Deepto Chakrabarty}
\affiliation{Department of Physics, Massachusetts Institute of Technology, Cambridge, MA 02139, USA}
\affiliation{Kavli Institute for Astrophysics and Space Research, Massachusetts Institute of Technology, Cambridge, MA 02139, USA}

\author[0000-0003-4780-4105]{Emma Chickles}
\affiliation{Department of Physics, Massachusetts Institute of Technology, Cambridge, MA 02139, USA}
\affiliation{Kavli Institute for Astrophysics and Space Research, Massachusetts Institute of Technology, Cambridge, MA 02139, USA}

\author[0000-0002-3168-0139]{Matthew J. Graham}
\affiliation{Division of Physics, Mathematics and Astronomy, California Institute of Technology, Pasadena, CA 91125, USA}

\author[0000-0003-0172-0854]{Erin Kara}
\affiliation{Department of Physics, Massachusetts Institute of Technology, Cambridge, MA 02139, USA}
\affiliation{Kavli Institute for Astrophysics and Space Research, Massachusetts Institute of Technology, Cambridge, MA 02139, USA}

\author[0000-0001-5390-8563]{S. R. Kulkarni}
\affiliation{Division of Physics, Mathematics and Astronomy, California Institute of Technology, Pasadena, CA 91125, USA}

\author[0000-0002-2498-7589]{Thomas R. Marsh}
\affiliation{Department of Physics, University of Warwick, Coventry CV4 7AL, UK}

\author[0000-0002-3310-1946]{Melania Nynka}
\affiliation{Kavli Institute for Astrophysics and Space Research, Massachusetts Institute of Technology, Cambridge, MA 02139, USA}

\author[0000-0002-8850-3627]{Thomas A. Prince}
\affiliation{Division of Physics, Mathematics and Astronomy, California Institute of Technology, Pasadena, CA 91125, USA}

\author[0000-0003-3769-9559]{Robert A. Simcoe}
\affiliation{Department of Physics, Massachusetts Institute of Technology, Cambridge, MA 02139, USA}
\affiliation{Kavli Institute for Astrophysics and Space Research, Massachusetts Institute of Technology, Cambridge, MA 02139, USA}

\author[0000-0002-2626-2872]{Jan van~Roestel}
\affiliation{Anton Pannekoek Institute for Astronomy, University of Amsterdam, 1090 GE Amsterdam, NL}
\affiliation{Division of Physics, Mathematics and Astronomy, California Institute of Technology, Pasadena, CA 91125, USA}

\author[0000-0002-0853-3464]{Zach Vanderbosch}
\affiliation{Division of Physics, Mathematics and Astronomy, California Institute of Technology, Pasadena, CA 91125, USA}

\author[0000-0001-8018-5348]{Eric C. Bellm}
\affiliation{DIRAC Institute, Department of Astronomy, University of Washington, 3910 15th Avenue NE, Seattle, WA 98195, USA}

\author[0000-0002-5884-7867]{Richard G. Dekany}
\affiliation{Caltech Optical Observatories, California Institute of Technology, Pasadena, CA, USA}

\author{Andrew J. Drake}
\affiliation{Division of Physics, Mathematics and Astronomy, California Institute of Technology, Pasadena, CA 91125, USA}

\author[0000-0003-3367-3415]{George Helou}
\affiliation{IPAC, California Institute of Technology, 1200 E. California Blvd, Pasadena, CA 91125, USA}

\author[0000-0002-8532-9395]{Frank J. Masci}
\affiliation{IPAC, California Institute of Technology, 1200 E. California Blvd, Pasadena, CA 91125, USA}

\author{Jennifer Milburn}
\affiliation{Caltech Optical Observatories, California Institute of Technology, Pasadena, CA, USA}

\author[0000-0002-0387-370X]{Reed Riddle}
\affiliation{Caltech Optical Observatories, California Institute of Technology, Pasadena, CA, USA}

\author[0000-0001-7648-4142]{Ben Rusholme}
\affiliation{IPAC, California Institute of Technology, 1200 E. California Blvd, Pasadena, CA 91125, USA}

\author[0000-0001-7062-9726]{Roger Smith}
\affiliation{Caltech Optical Observatories, California Institute of Technology, Pasadena, CA, USA}



\begin{abstract}
We report the discovery of ZTF J0127+5258, a compact mass-transferring binary with an orbital period of 13.7 minutes. The system contains a white dwarf accretor, which likely originated as a post-common envelope carbon-oxygen (CO) white dwarf, and a warm donor ($T_{\rm eff,\,donor}= 16,400\pm1000\,\rm K$). The donor probably formed during a common envelope phase between the CO white dwarf and an evolving giant which left behind a helium star or helium white dwarf in a close orbit with the CO white dwarf. We measure gravitational wave-driven orbital inspiral with $\sim 35\sigma$ significance, which yields a joint constraint on the component masses and mass transfer rate. While the accretion disk in the system is dominated by ionized helium emission, the donor exhibits a mixture of hydrogen and helium absorption lines. Phase-resolved spectroscopy yields a donor radial-velocity semi-amplitude of $771\pm27\,\rm km\, s^{-1}$, and high-speed photometry reveals that the system is eclipsing. We detect a {\it Chandra} X-ray counterpart with $L_{X}\sim 3\times 10^{31}\,\rm erg\,s^{-1}$. Depending on the mass-transfer rate, the system will likely evolve into either a stably mass-transferring helium CV, merge to become an R Crb star, or explode as a Type Ia supernova in the next million years. We predict that the Laser Space Interferometer Antenna (LISA) will detect the source with a signal-to-noise ratio of $24\pm6$ after 4 years of observations. The system is the first \emph{LISA}-loud mass-transferring binary with an intrinsically luminous donor, a class of sources that provide the opportunity to leverage the synergy between optical and infrared time domain surveys, X-ray facilities, and gravitational-wave observatories to probe general relativity, accretion physics, and binary evolution.

\end{abstract}



\section{Introduction} \label{sec:intro}

Our understanding of how the shortest orbital period binary systems in the Galaxy form is undergoing a renaissance due to discoveries facilitated by optical time domain surveys such as the Zwicky Transient Facility (ZTF) \citep{Masci2019,Graham2019,Dekany2020,Bellm2019}. We have known for decades that there exists a substantial population of helium-rich cataclysmic variables (helium CVs, also known as AM CVns) with orbital periods under the canonical period minimum of 80 minutes of hydrogen-rich cataclysmic variables. About a hundred of these systems are currently known \citep{Ramsay2018}, and they have a few defining characteristics:

\begin{itemize}
        \item Systems with orbital periods below about 30 minutes exhibit high-state accretion disks \citep{Duffy2021}, which consistently maintain a high rate of mass transfer, whereas systems with longer orbital periods (30-70 minutes) have quiescent phases, and undergo irregular outbursts with a recurrence timescale that lengthens at longer orbital periods \citep{Levitan2015}. Until recently, these longer-period systems, which are most easily detected via outbursts, have dominated the known population of ultracompact binaries. 
    \item The donors in these systems remain largely undetected, except  when they transit the accreting white dwarf \citep{vanRoestel2022}. This is because the donors are quite cool, likely because they have undergone adiabatic expansion as a result of mass loss after reaching a minimum orbital period. Thus, it is expected that most of the known population of helium CVs, which host cold donors, are post-period-minimum objects.
\end{itemize}

There are three main proposed progenitor channels for helium CVs: the interaction of double white dwarfs \citep{Marsh2004}, the interaction of a white dwarf accretor with a helium-burning star \citep{Kupfer2020a}, and transitional cataclysmic variables, in which a white dwarf interacts with an evolved main-sequence donor. The discovery of ZTF J1813+4251, a 51-minute orbital period transitional CV with an F-type donor \citep{Burdge2022b} cemented the role of the transitional CV channel in producing at least some of these systems. Several mass-transferring white dwarf plus helium star pairs have been discovered in recent years \citep{Burdge2020a,Kupfer2020a,Kupfer2020b,Kupfer2022}, though it is unclear whether these systems will evolve into the post-period-minimum helium CVs we see, or undergo a thermonuclear supernova well before reaching such orbital periods. Finally, we know of many detached short-orbital-period double-white-dwarf systems, the shortest of which are the eclipsing binaries ZTF J1539+5027 \citep{Burdge2019a} and ZTF J2243+5242 \citep{Burdge2020b} (with periods of $\sim6.9$ and $\sim8.8$ minutes, respectively), and at least two interacting white dwarf systems which are undergoing direct impact accretion \citep{Marsh2004}, HM Cancri \citep{Ramsay2002} and V407 Vul \citep{Marsh2002} (with orbital periods of $\sim5.4$ and $\sim9.5$ minutes, respectively). HM Cancri hosts a carbon-oxygen core (CO) white dwarf accretor, with a likely helium core white dwarf donor, and V407 Vul likely consists of the same configuration, though it is difficult to study the characteristics of these systems because HM Cancri's flux is dominated by the accretor (with some hint of reprocessed radiation from the donor), while V407 Vul's optical flux output is dwarfed by a spatially coincident G type main sequence star (it is unclear whether it is associated with the ultracompact binary). Both systems exhibit a negative period derivative, indicating that they are likely pre-period minimum; however, recent work has suggested that HM Cancri is exhibiting a slowing in its orbital decay rate, and may reach a period minimum in the near future \citep{Strohmayer2021,Munday2023}. There is ongoing debate whether these examples of double white dwarfs indeed contribute to the population of post-period minimum stably mass transferring helium CVs which have been observed over the last few decades, as \citet{Shen2015} suggest that even systems with low mass ratios in which mass-transfer would normally be stable are driven towards merger due to nova shells expelled by the accretor interacting with the donor. Additionally, \citet{Brown2016} found that observationally, the space density of close CO+He core WD binaries exceeds the population of stably mass-transferring helium CVs by a factor of 40, suggesting that the vast majority of double white dwarfs merge, or that we are significantly underestimating the space density of helium CVs, most of which have been selected on the basis of outbursting behavior.

Here, we report the discovery and characterization of ZTF J0127+5258, an accreting binary system with an orbital period of $13.7$ minutes. Remarkably, the system hosts a luminous donor which is clearly visible in the optical, contributing comparable luminosity to the high state accretion disk surrounding its white dwarf companion. The donor exhibits primarily helium lines, but still manifests hydrogen absorption as well, an atmosphere very similar to the pre-mass transfer 20-minute orbital period ellipsoidal variable PTF J0533+0209 \citep{Burdge2019b}. We detect the presence of an accretion disk, which manifests itself both in the form of prominent He II emission in the optical spectrum, as well as in a broad triangular-shaped eclipse of the warm donor star. Using archival photometric observations, we detect that the system is undergoing rapid orbital decay, suggesting that it has not yet reached a period minimum. Because of the elevated temperature of its donor, the system likely did not originate from the continuously mass-transferring transitional CV channel described in \citet{Burdge2022b}, but instead probably represents the final inspiral phase of either the double white dwarf channel, in which an ELM white dwarf recently overflowed its Roche lobe, or the helium star channel, though we disfavor this scenario due to the presence of hydrogen at the surface of the donor. In this letter, we discuss the observational properties of this system and its implications for compact binary evolution. 

\section{Observations} \label{sec:observations}

\subsection{Discovery in ZTF data}
We discovered ZTF J0127+5258 using archival photometry from the Zwicky Transient Facility, as part of an ongoing campaign to identify short orbital period binaries \citep{Burdge2020a}. The object was discovered in a search for periodic photometric variations in 20 million objects selected as being underluminous relative to the main sequence of stars in \emph{Gaia} DR2 data \citep{Gaia2018}; this selection and the period search are described in further detail in \citet{Burdge2022a,Burdge2020a}. We note that ZTF J0127+5258 falls in a chip gap in ZTF's primary field grid, and thus only contains photometry from the secondary field grid. This grid has much poorer sampling than the primary field grid, with ZTF J0127+5258 averaging less than 90 epochs per year in the ZTF g and r filters combined, demonstrating the viability of discovering ultracomapct binaries at even relatively sparse cadences.

\subsection{High-speed photometry}

After discovering the source in ZTF data, we observed ZTF J0127+5258 using the Caltech HIgh-speed Multi-color camERA (CHIMERA) \citep{Harding2016} on the 200-inch Hale telescope at Palomar observatory. We obtained these observations on four separate occasions, in the $g^{\prime}$ filter on the blue channel of the instrument, and the $r^{\prime}$ and $i^{\prime}$ filters on the red channel of the instrument. These observations took place on July 23, 2020, November 14, 2020, Oct 10, 2021, and Oct 31, 2021. The absolute timestamps of exposures were time-tagged to better than millisecond accuracy using a GPS receiver.

\subsection{Spectroscopic observations}

On October 15, 2020, we obtained phase-resolved spectroscopic observations of ZTF J0127+5258 using the Low-Resolution Imaging Spectrometer (LRIS) \citep{Oke1995} on the 10-m W.\ M.\ Keck I Telescope on Mauna Kea. We adopted a 2-minute exposure time and obtained a total of 39 exposures over approximately 96 minutes, covering 7 orbital cycles of the system. We binned these exposures $4\times4$ in the blue channel and $2\times2$ in the red channel to minimize readout time and readout noise and used the 600/4000 grism as the dispersive element on the blue channel, and the 400/8500 grating as the dispersive element for the red channel.

\section{Analysis and interpretation} \label{sec:analysis}

\subsection{Lightcurve}

\begin{figure}
\includegraphics[width=0.46\textwidth]{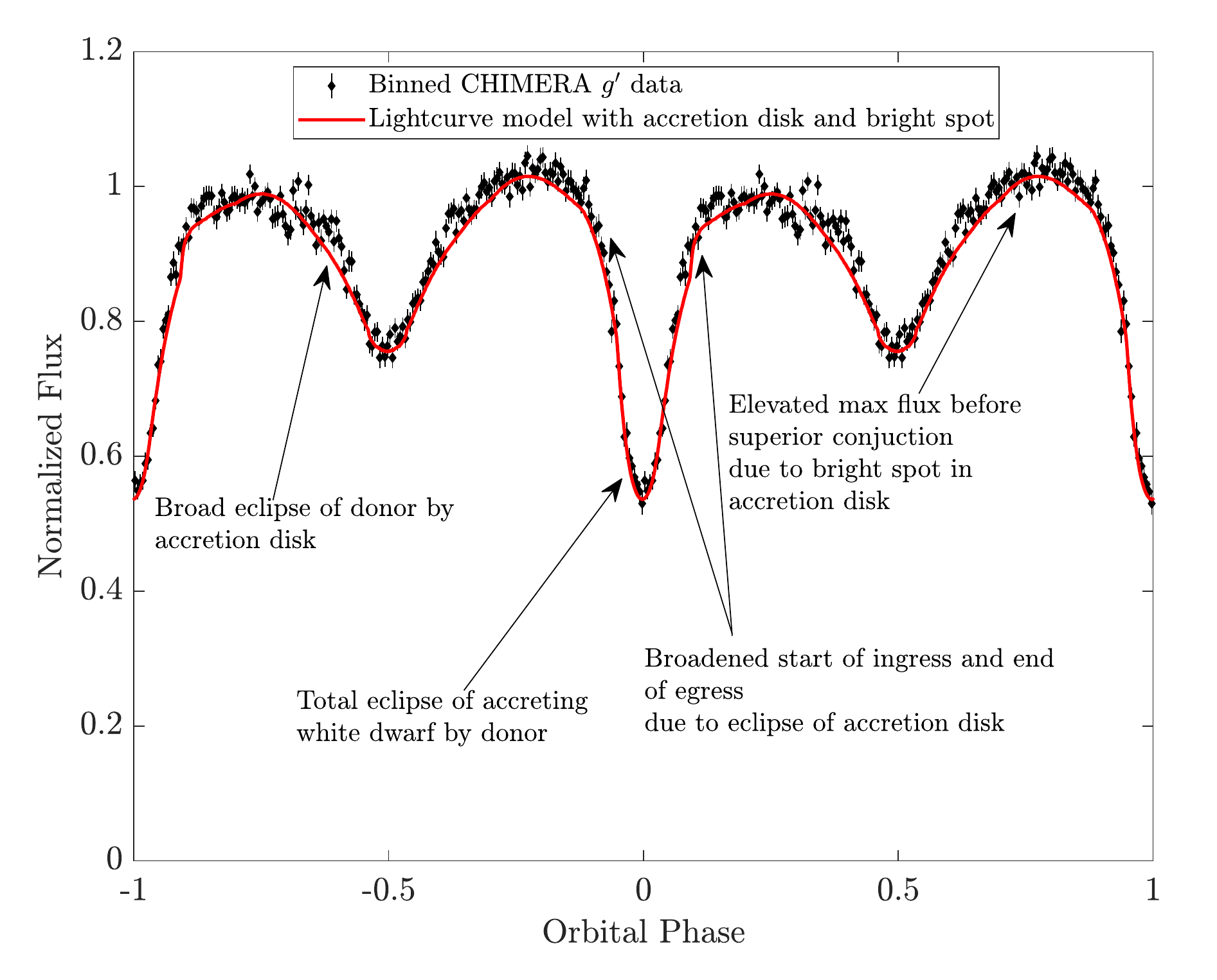}
\caption{A model fit of the binned CHIMERA $g^{\prime}$ lightcurve of ZTF J0127+5258. We used the LCURVE modeling code to construct a physical model for the system. In order to account for the behavior in the lightcurve, particularly the geometry of the eclipses, as well as the increased flux at the quadrature phase at $0.75$, we had to include an accretion disk with a bright spot, indicating that the disk and bright spot contribute an appreciable fraction of the optical flux. 
\label{fig:LC}}
\end{figure}

Figure \ref{fig:LC} shows the CHIMERA $g^\prime$ light curve of ZTF J0127+5258 and a representative model light curve. 
We used the LCURVE modeling code \citep{Copperwheat2010} because in modeling the lightcurve, we discovered that we could only account for the complex structure by including an accretion disk and a bright impact spot of the accretion stream onto the outer edge of the disk.

We do not present parameter estimates for the component radii and masses based on the example lightcurve model, because including an accretion disk and bright spot introduces a significant number of free parameters, many of which are degenerate with physical parameters of interest. The model illustrated in Figure \ref{fig:LC} is based on a plausible set of parameters, with the accreting white dwarf radius fixed to that expected for a $\sim 0.75 \rm M_{\odot}$ WD, a mass ratio of $\sim 0.2$, a donor which is Roche-lobe filling, and an accretion disk with an inner edge at the radius of the accreting WD, and outer edge at the radius of a hot spot, fixed to approximately 70 percent the Roche lobe radius of the accretor based on the maximum radius of a stable accretion disk \cite{Paczynski1971}. In the future, a higher signal-to-noise lightcurve obtained with an instrument such as HiPERCAM may allow for a clear delineation of the ingress/egress structure in the eclipse arising from the two bodies transiting each other from structure that originates from the accretion disk and hot spot eclipse; if the substructure of the eclipse can be characterized, then this system could be robustly constrained on the basis of Kepler's laws and Roche geometry alone, like the case of ZTF J1813+4251 \citep{Burdge2022b}. 

\subsection{Orbital inspiral}

\begin{figure}
\includegraphics[width=0.46\textwidth]{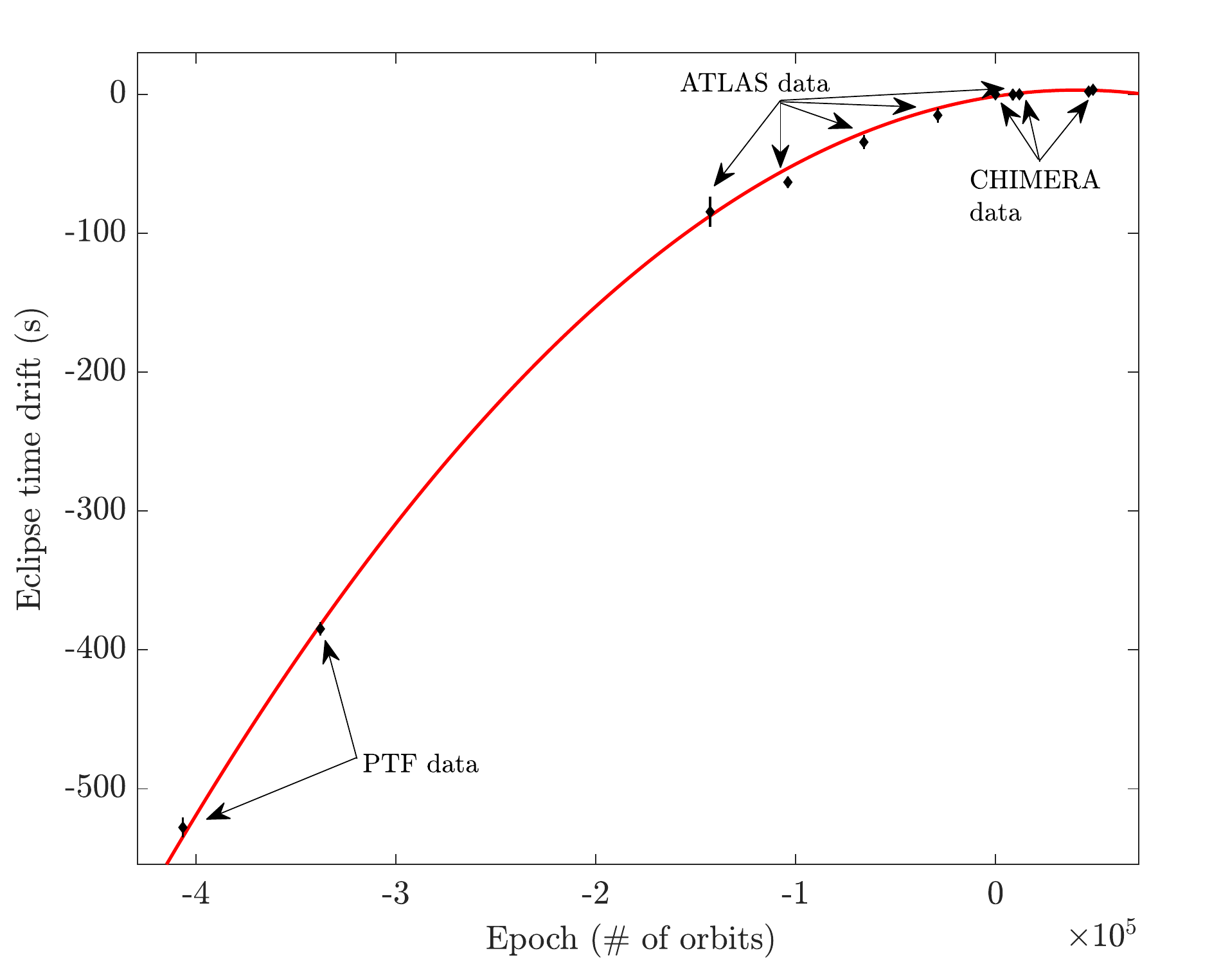}
\caption{The measured evolution of eclipse times in ZTF J0127+5258. The red fit of a quadratic clearly indicates that the system is undergoing orbital decay, as the coefficient of the quadratic term of the polynomial is clearly negative. A precise measurement of the orbital evolution was made possible by the eclipsing nature of the source (allowing for precise timing), and the long baseline resulting from archival PTF and ATLAS observations.
\label{fig:Decay}}
\end{figure}

After constructing the lightcurve model illustrated in Figure \ref{fig:LC}, we fit this model to archival PTF \citep{Law2009} and ATLAS \citep{Tonry2018,Heinze2018} lightcurves, as well as our four CHIMERA epochs, in order to investigate whether any measurable orbital evolution was occurring in the system--see \citet{Burdge2019a} for further details on this procedure. As illustrated by the polynomial fit to the data in Figure \ref{fig:Decay}, we clearly detect orbital decay is occurring, at a rate of $-6.53\pm0.19\times 10^{-12}\,\rm s\, s^{-1}$. Because ZTF J1027+5258 is mass transferring, translating this period decay into a direct constraint on the component masses is nontrivial; we discuss this further in Section~\ref{sec:mdot_pdot}. Nevertheless, because of the orbital decay and warm donor, we can deduce that the system is almost certainly pre-period minimum, and in the final phase of its inspiral. 

\subsection{Spectroscopy}

\begin{figure}
\includegraphics[width=0.46\textwidth]{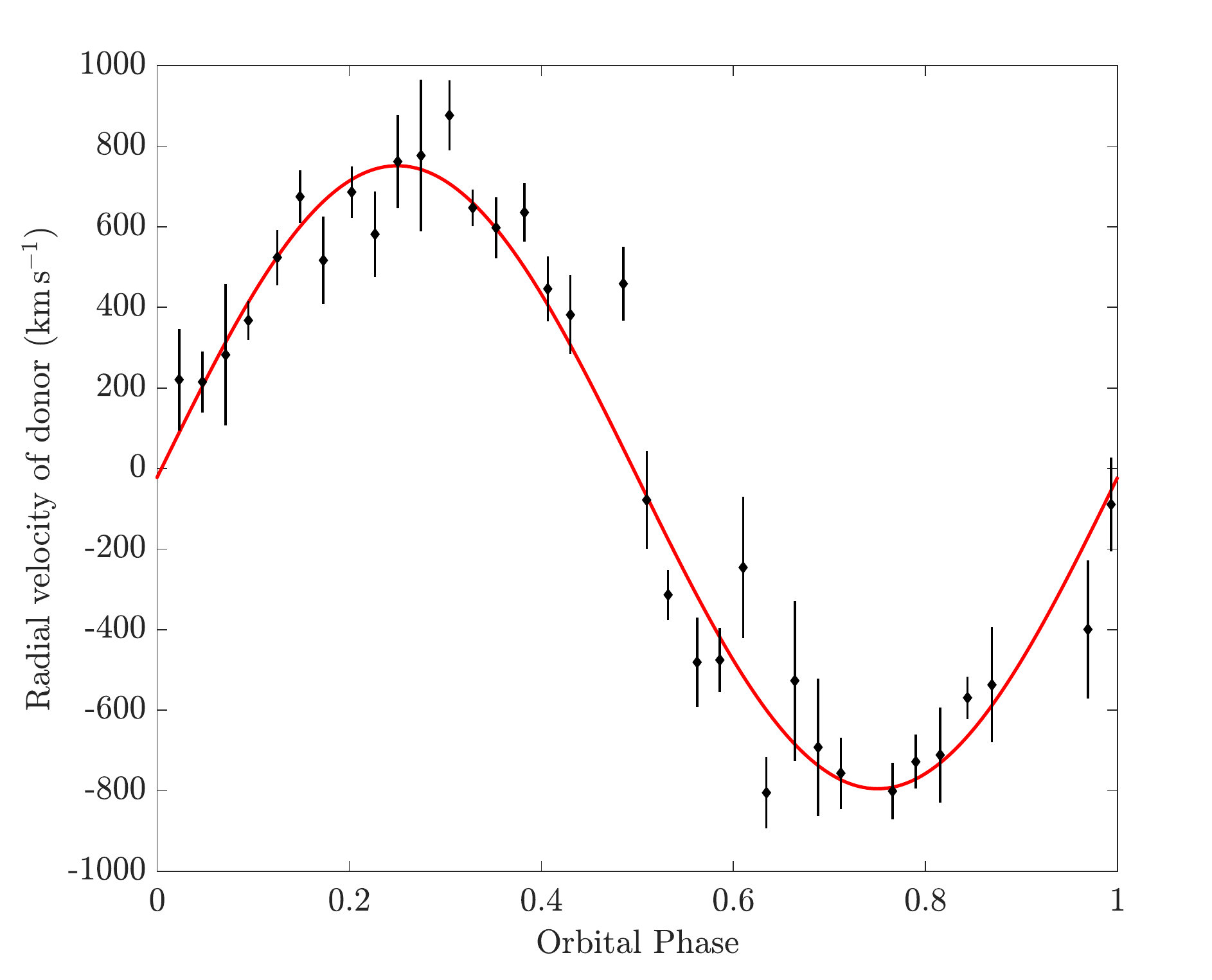}
\caption{The measured radial velocities of the hydrogen absorption lines in the spectrum of ZTF J0127+5258. The orbital phases are referenced to those determined from the mid-eclipse time in the lightcurve model. The poor fit near inferior conjunction (phase $0.5$) may originate from a Rossiter–McLaughlin effect as the accretion disk and accreting white dwarf transit the donor.
\label{fig:RVs}}
\end{figure}

\begin{figure}
\includegraphics[width=0.46\textwidth]{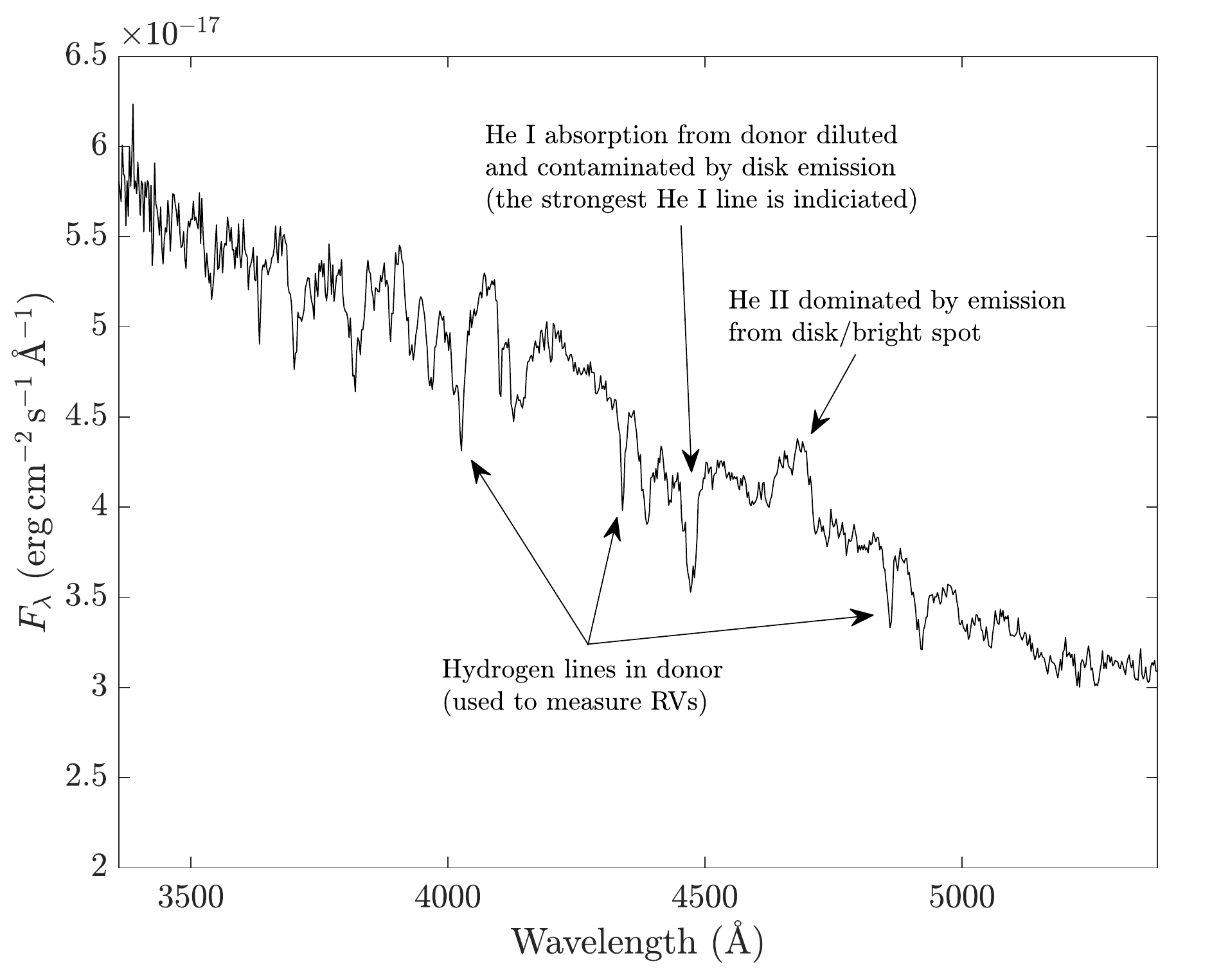}
\caption{The extinction-corrected optical spectrum of ZTF J0127+5258 obtained with LRIS on the Keck I telescope. Much like the spectrum of PTF J0533+0209 \citep{Burdge2019b}, the atmosphere of the donor exhibits a mixture of hydrogen and helium absorption lines, indicating that it is helium-rich but still contains traces of hydrogen. Because of the contribution of the accretion disk to both the continuum and the dilution of the absorption lines by emission, we do not fit a model atmosphere to the object. Instead, we use the spectral energy distribution to estimate the temperature of the object.
\label{fig:spectrum}}
\end{figure}

In order to measure the radial velocity semi-amplitude of the donor, we fit Voigt profiles to the Balmer series of hydrogen absorption lines, which appear to originate primarily from the donor's atmosphere. Figure \ref{fig:RVs} illustrates the radial velocity measurements resulting from this analysis, which are consistent with a semi-amplitude of $771\pm27\,\rm km\, s^{-1}$. We also tried to fit the numerous helium I absorption lines in the spectrum, and found a smaller semi-amplitude, with a phase strongly inconsistent with the mid-eclipse time (whereas the radial velocities from the Balmer lines yielded a phase in good agreement with the eclipse time in the lightcurve). We attribute the strange behavior of the helium I lines to contamination and dilution by emission originating from the accretion disk/bright spot. This accretion-associated emission manifests itself even more strongly in the helium II lines, as indicated in Figure \ref{fig:spectrum}.

\subsection{SWIFT Observation}

We observed ZTF J0127+5258 using the XRT and UVOT instruments aboard the Swift observatory with a 3 kilosecond exposure time (observation ID: 14103001). We do not detect the source in the XRT observations, however, we see an ultraviolet counterpart in the UVOT image (taken in the UVW2 filter). We obtain a source brightness of $21.74\pm 0.15\,\mathrm{(stat)}\,\pm 0.03\,\mathrm{(sys)}\, m_{AB}$ in this filter, which we estimate corresponds to a de-reddened apparent magnitude of approximately $19.37\, \rm m_{AB}$. However, the uncertainties are large in the ultraviolet luminosity of the source, due to the large degree of sensitivity of ultraviolet flux to extinction corrections.

\subsection{SED analysis and distance/temperature estimate}

Because of its faint apparent magnitude, ZTF J0217+5258 lacks a well-constrained parallax in \emph{Gaia} DR3, with a reported value of $\pi=0.44\pm0.47$. If we treat the Gaia parallax, $\pi$, as having a Gaussian probability distribution with a most probable value of 0.44 mas and an rms scatter of 0.47 mas, at least for $\pi > 0$, then the corresponding distance distribution has a maximum at 1100 pc, and $10\%$ limits of 600 pc on the close side and 4200 pc on the distant side. In addition to these basic constraints we can infer from the parallax, we also explore alternative approaches to estimating the distance to the source. Our first approach is based on Figure 3 of \citet{Ramsay2018}, which illustrates the absolute magnitude of various accreting ultracompact binaries as a function of the orbital period. The dynamic range of luminosity in this figure is large, especially at the short orbital periods relevant to ZTF J0127+5258, but we estimate that the absolute magnitude of the source should fall in the range of 5-8 in \emph{Gaia} G given its orbital period. We compute the posterior distribution of our distance estimates by assuming a uniform prior in the absolute magnitude of the source in the range of 5-8, applying an extinction correction of $-0.75$ magnitudes to the \emph{Gaia} G filter based on the observed reddening of $E(g-r)=0.29$ at distances of $>3 \rm \, kpc$ \cite{Green2019}. Based on its apparent magnitude and this reddening and the typical absolute magnitude of interacting ultracompact binaries in this period range, we estimate that the source is approximately $3.5^{+1.7}_{-1.5}\,\rm kpc$ from the Solar System.

To obtain a more precise distance estimate (and also a temperature estimate), we fit the spectral energy distribution (SED) of the source. Although the source is reddened, the gradient of the reddening is small at distances $>1.5\,\rm kpc$, and thus we fix the reddening to $E(g-r)=0.29$. We use the Kernel density estimator Ultranest \citep{Buchner2014,Buchner2017,Buchner2021} to fit the SED photometry of the source, including the Swift UVOT UVW2 measurement. In modeling the SED, we sample over the masses of the accretor, which accounts for luminosity contributions from the white dwarf and the disk, and donor (giving both a uniform prior of $0-1.4\, M_{\odot}$), fix the radius of the donor to its Roche lobe, and allow the effective radius of the accretor to vary between 0 up to $95$ percent of its Roche lobe (a limit set by the instability criterion for accretion disks). Using our CHIMERA lightcurve, we also determine that on average, the donor contributes 50 percent of the flux in g-band, and thus require that the ratio of flux contributed by the two components in this filter be confined within the range of $1\pm0.1$. We sample over effective temperatures for both the donor and accretor. We find an effective donor temperature of $16400^{+1000}_{-900}\, \rm K$, and a distance of $3.1^{+0.5}_{-0.6}\,\rm kpc$, consistent with the distance predicted based on the luminosity of other accreting binaries at similar orbital periods.

\subsection{Chandra Observation}

We observed ZTF J0127+5258 using the ACIS-I instrument aboard the Chandra X-ray observatory (Proposal ID: 23300558) for $16362\, \rm s$. In this exposure, we recorded 26 photons within $2\arcsec$, one of which we omitted as likely a background photon due to its high energy of $\sim 13\, \rm keV$. We binned the photons into five spectral bins, and used the XSPEC tool to fit for the flux and photon index of a power law model of the spectrum. We fixed the column density to $n_H=1.6\times 10^{21}$ based on the estimated reddening of $\mathrm(E(g-r))=0.29$. We obtain an unabsorbed flux value of $3.0^{+0.8}_{-0.6}\times10^{-14}\,\rm erg\,cm^{-2}\,s^{-1}$ in the $0.2-8.0\,\rm keV$ bandpass, and a power law index of $\gamma=1.33\pm0.44$. Given the source distance and the column density, we estimate that the X-ray luminosity of the source in the $0.2-8.0\,\rm keV$ band is $3.5\pm1.3\times 10^{31}\, \rm erg \, s^{-1}$. This luminosity is comparable to the $10.4\rm \, min$ accreting AM CVn binary ES Ceti, which exhibits an X-ray luminosity of approximately $8.3 \times 10^{30}\, \rm erg \, s^{-1}$ in the $0.2-5.0\,\rm keV$ band \citep{Strohmayer2004}. We note that because of the edge-on inclination of the system, this X-ray flux represents only a fraction of the bolometric X-ray luminosity of the system, and likely arises from the boundary layer of the accretion disk, which is at least partially obscured at near edge-on inclinations.

\begin{deluxetable}{cl}[htbp]
\tablenum{1}
\tablecaption{Measured parameters \label{tab:Parameters}}
\tablehead{\colhead{Quantity:} & \colhead{Measured value}}
\tablewidth{1000pt}

\startdata
$\rm RA$ & 01:27:47.62 (h:m:s)    \\
$\rm Dec$ & 52:58:13.03 (d:m:s)    \\
$\pi$ & $0.44\pm0.47\,\rm mas\, yr^{-1}$    \\
$\mu_{\rm RA}$ & $-1.14\pm0.41\,\rm mas\, yr^{-1}$    \\
$\mu_{\rm Dec}$ & $-2.57\pm0.42\,\rm mas\, yr^{-1}$    \\
$\rm G_{mag}$ & $19.81\pm0.01\,\rm (Vega\,\, magnitude)$    \\
\hline
$K_{\mathrm{Donor}}$ & $771\pm27\,\rm km\, s^{-1}$    \\
$T_0 $ & $59518.2991305\pm0.0000081\,\rm MBJD_{TDB}$     \\
$P_b $ & $822.68031456\pm0.000043\, \rm s$     \\
$\dot{P}_b$ & $-6.53\pm0.19\times 10^{-12}\,\rm s\, s^{-1}$    \\
$T_{\mathrm{Donor}}$ & $16400^{+1000}_{-900}\, \rm K$    \\
\enddata
\tablecomments{Measured parameters for ZTF J0127+5258. Coordinates are given at epoch J2000.0, using J2000 equinox.}
\end{deluxetable}

\begin{deluxetable*}{ccl}[htbp]
\tablenum{2}
\tablecaption{Inferred component parameters \label{tab:inferred}}
\tablehead{\colhead{Quantity:} & \colhead{$\log(\dot{M}/(M_{\odot}\,\rm yr^{-1}))\sim \mathcal{N}(-8.5, 0.5)$} & \colhead{$\log(\dot{M}/(M_{\odot}\,\rm yr^{-1}))\sim \mathcal{N}(-7.3, 0.5)$}}
\tablewidth{100pt}

\startdata
$M_{\mathrm{accretor}} $ & $0.75\pm0.06\,M_{\odot}$ & $0.87\pm0.11\,M_{\odot}$   \\
$M_{\mathrm{donor}} $ & $0.19\pm0.03\,M_{\odot}$ & $0.31\pm0.11\,M_{\odot}$   \\
$R_{\mathrm{donor}} $ & $0.051\pm0.003\,R_{\odot}$ & $0.059\pm0.008\,R_{\odot}$   \\
$L_{\mathrm{donor}} $ & $0.17\pm0.05\,L_{\odot}$ & $0.23\pm0.08\,L_{\odot}$   \\
\enddata
\tablecomments{Inferred component parameters for ZTF J0127+5258 based on the measured parameters. Because of the uncertainty in the mass transfer rate and its implications for the parameter estimates, we report two columns illustrating how the inferred parameters change assuming the high vs low mass transfer rate cases. }
\end{deluxetable*}

\subsection{Effect of mass transfer on period decay}
\label{sec:mdot_pdot}
In detached binaries, a measurement of the period derivative yields a direct constraint on the chirp mass of the system. Mass transfer complicates this constraint since the immediate effect of transferring mass from the lower-mass component to the higher-mass component is widening of the orbit. This can be seen by considering the orbital angular momentum of a binary with a circular orbit:
\begin{equation}
    \label{eq:Jorb}
    J_{{\rm orb}}=\frac{M_{{\rm wd}}M_{{\rm donor}}}{\left(M_{{\rm wd}}+M_{{\rm donor}}\right)^{1/3}}P_{{\rm orb}}^{1/3}\left(\frac{G^{2}}{2\pi}\right)^{1/3}.
\end{equation}
Solving for $P_{\rm orb}$ and taking a logarithmic derivative yields: 
\begin{equation}
    \frac{\dot{P}_{{\rm orb}}}{P_{{\rm orb}}}=3\frac{\dot{J}_{{\rm orb}}}{J_{{\rm orb}}}+\frac{\dot{M}_{{\rm tot}}}{M_{{\rm tot}}}-3\frac{\dot{M}_{{\rm wd}}}{M_{{\rm wd}}}-3\frac{\dot{M}_{{\rm donor}}}{M_{{\rm donor}}}, 
\end{equation}
where $M_{\rm tot} = M_{\rm wd} + M_{\rm donor}$. For conservative mass transfer, we expect $\dot{M}_{{\rm tot}}=0$ and $\dot{M}_{\rm wd} = -\dot{M}_{\rm donor}$. If we then define $\dot{M}\equiv \dot{M}_{\rm wd}$ (a positive quantity), we obtain
\begin{equation}
    \label{eq:pdot}
    \frac{\dot{P}_{{\rm orb}}}{P_{{\rm orb}}}=3\frac{\dot{J}_{\rm orb}}{J_{\rm orb}}+3\frac{\dot{M}}{M_{\rm donor}}\left(1-q\right),
\end{equation}
where $q=M_{\rm donor}/M_{\rm wd}$. Assuming the only significant angular momentum loss comes from gravitational waves, we have 
\begin{equation}
    \frac{\dot{J}_{{\rm orb}}}{J_{{\rm orb}}}=-\frac{32G^{5/3}}{5c^{5}}\frac{M_{{\rm wd}}M_{{\rm donor}}}{\left(M_{{\rm wd}}+M_{{\rm donor}}\right)^{1/3}}\left(\frac{2\pi}{P_{{\rm orb}}}\right)^{8/3}.
\end{equation}

To constrain the masses of both components in the presence of mass transfer, we fit the observed period derivative and donor RV semi-amplitude, while leaving the component masses, inclination, and mass transfer rate as free parameters. The likelihood function compares the observed and predicted semi-amplitudes and period derivatives:

\begin{equation}
\begin{split}
    \ln L=-\frac{1}{2}\frac{\left(K_{{\rm donor,\,pred}}-K_{{\rm donor,\,obs}}\right)^{2}}{\sigma_{K_{{\rm donor}}}^{2}}\\-\frac{1}{2}\frac{\left(\dot{P}_{{\rm orb,\,pred}}-\dot{P}_{{\rm orb,\,obs}}\right)^{2}}{\sigma_{\dot{P}_{{\rm orb}}}^{2}},
\end{split}
\end{equation}
where $\dot{P}_{\rm orb,\,pred}$ is the period derivative predicted by Equation~\ref{eq:pdot} (assuming mass transfer is conservative), and $K_{{\rm donor,\,pred}}=\left[\frac{2\pi G}{\left(M_{{\rm donor}}+M_{{\rm wd}}\right)^{2}P_{{\rm orb}}}\right]^{1/3}M_{{\rm wd}}\sin i$, with $i$ the inclination. 

Motivated by the models of \citet{Wong2021}, we adopt two different priors on the mass transfer rate. The first is a relatively low value, $\log(\dot{M}/(M_{\odot}\,\rm yr^{-1}))\sim \mathcal{N}(-8.5, 0.5)$ (this notation indicates a normal distribution centered at $-8.5$ with standard deviation $0.5$, which matches their models for a low-mass white dwarf donor. However, these models only undergo Roche-lobe overflow at shorter periods and are already evolving toward longer periods by the time they reach 13.7 minutes. Their models in which the donor is a helium star begin mass transfer at longer periods ($\sim 20$ minutes) and thus do pass through 13.7 minutes while the period is decreasing \citep[see also][]{Brooks2015}. These models have significantly higher mass transfer rates, and for these, we adopt $\log(\dot{M}/(M_{\odot}\,\rm yr^{-1}))\sim \mathcal{N}(-7.3, 0.5)$.

\begin{figure*}
    \centering
    \includegraphics[width=\textwidth]{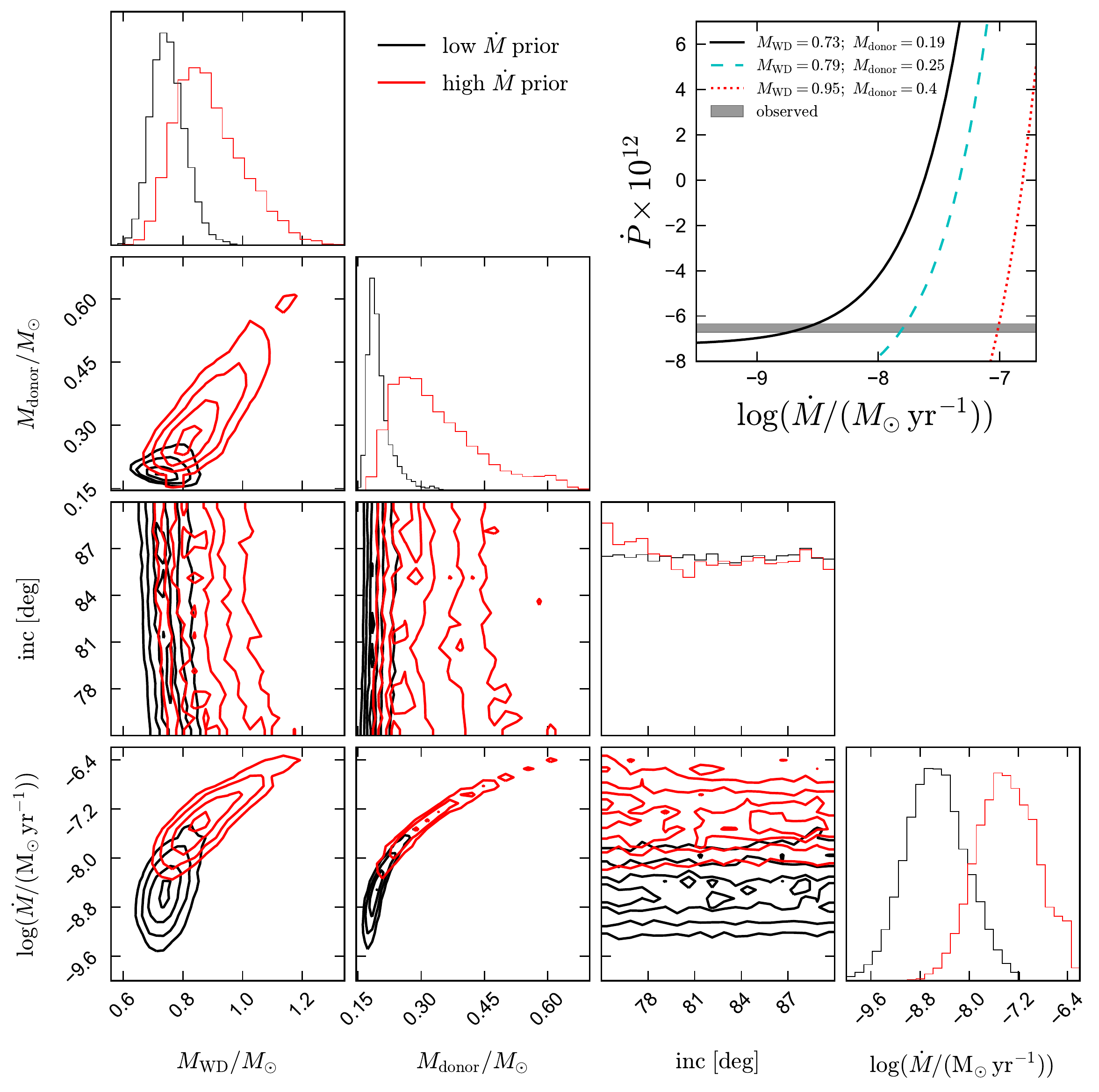}
    \caption{Constraints from joint fitting of the donor's RV semi-amplitude and the observed period derivative. Black and red contours adopt priors of $\dot{M}\sim \mathcal{N}(-8.5, 0.5)$ and $\dot{M}\sim \mathcal{N}(-7.3, 0.5)$, respectively. The upper right inset shows how, for two different combinations of $M_{\rm WD}$ and $M_{\rm donor}$ that match the observed RV amplitude, the predicted period derivative depends on the mass transfer rate. Mass transfer widens the orbit, so a larger $\dot{M}$ requires higher masses to match the observed $\dot{P}_{\rm orb}$.  }
    \label{fig:corner}
\end{figure*}

Since the system is eclipsing, we use a prior on the inclination, $i\sim \mathcal{U}(75, 90)$. We use flat priors on $M_{\rm wd}$ and $M_{\rm donor}$. The results of this fitting are shown in Figure~\ref{fig:corner}.

In general, we infer higher masses for both components with the high-$\dot{M}$ prior. Since mass transfer widens the orbit and counteracts the GR-driven orbital inspiral, a larger value of $\dot{M}$ requires higher component masses to produce the same net period derivative. 

For the low-$\dot{M}$ prior, the orbit-widening effects of mass transfer are relatively modest for all values of $\dot{M}$ allowed by the prior, and as a result, the inferred component masses depend only weakly on $\dot{M}$. However, for the high-$\dot{M}$ prior, large enough values of $\dot{M}$ are supported that they can significantly change the expected $\dot{P}$, and so the inferred component masses have larger uncertainties. For values of $\dot{M}$ above $10^{-7}\,M_{\odot}\,\rm yr^{-1}$, the total mass required to reproduce the observed $\dot{P}$ and $K_{\rm donor}$ even exceeds the Chandrasekhar mass. 

An accurate measurement of the system luminosity can constrain $\dot{M}$. For a high state system like J0127, the expected accretion luminosity is $L_{{\rm acc}}\sim G\dot{M}M_{{\rm wd}}/\left(2R_{{\rm wd}}\right)\sim11\,L_{\odot}\left(\frac{\dot{M}}{10^{-8}\,M_{\odot}\,{\rm yr}^{-1}}\right)\left(\frac{M_{{\rm wd}}}{0.7\,M_{\odot}}\right)\left(\frac{R_{{\rm wd}}}{0.01\,R_{\odot}}\right)$. The high-$\dot{M}$ prior thus implies an accretion luminosity of $L_{\rm acc}\sim 50\,L_{\odot}$. The low-$\dot{M}$ prior implies  $L_{\rm acc}\lesssim 5\,L_{\odot}$. The presence of hydrogen in the donor spectrum suggests a recent onset of mass transfer, in better agreement with the evolutionary scenario leading to the low-$\dot{M}$ prior. Additionally, the low bolometric luminosity of the donor (see Table \ref{tab:inferred}) also favors the low mass transfer rate scenario, since the eclipse depth indicates roughly equal luminosity of the donor and the accretor in the optical--however, because most of the accretion luminosity is likely emitted in the ultraviolet, this observation cannot definitively rule out the high mass transfer rate case.

\subsection{He WD and He Star models}
\label{sec:MESA models}

We compare ZTF J0127+5258 to Modules for Experiments in Stellar Astrophysics \citep[$\mesa$;][]{MESAI,MESAII,MESAIII,MESAIV,MESAV,MESAVI}  models. The He WD models are made in a similar fashion to \cite{Wong2021}, assuming a non-rotating donor with orbital angular momentum loss due to gravitational wave emission. The initial models are made by stripping the envelope of an evolved star once a He core is formed, down to a hydrogen mass fraction of $X=0.1$, and allowing it to cool to the desired central specific entropy, $s$, with element diffusion enabled. 
For comparison, we also include a \mesa{} model with a $0.37\, \msun$ He star donor based on \cite{Bauer2021}, which can evolve through a similar orbital and $\Teff$ configuration as the highest entropy He WD donor models, but cannot satisfy the constraint of having some residual hydrogen at its surface by the time $\Porb = 13.7\,\mathrm{min}$. The He star model represents roughly a maximum radius/entropy configuration relative to a He WD model because the He star comes into contact while core He burning is ongoing, then evolves adiabatically once it donates enough mass to quench core He burning ($M \lesssim 0.3\,\msun$).

\begin{figure*}
    \centering
    \includegraphics[width=0.49\textwidth]{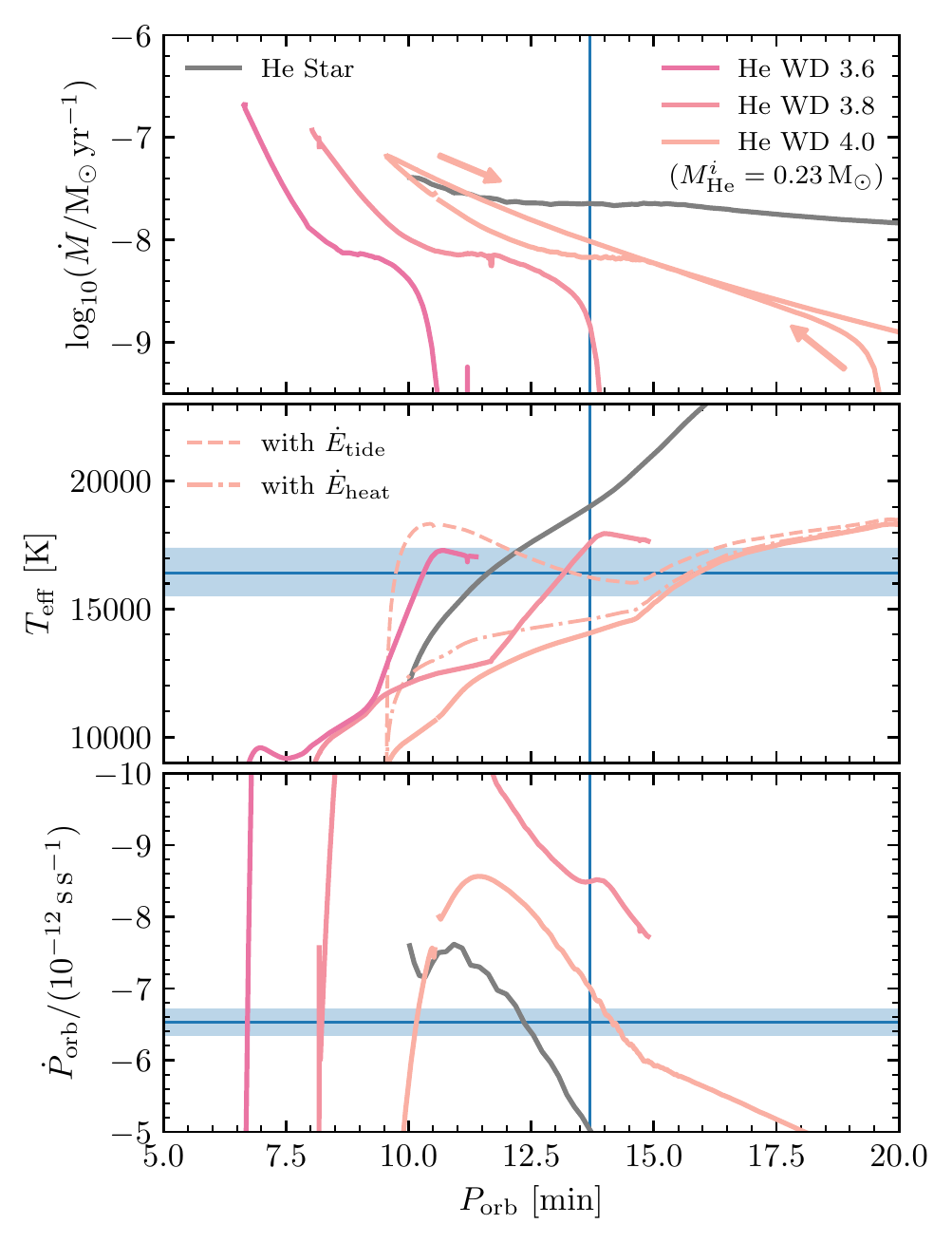}
    \caption{ Mass transfer rate (top), donor effective temperature (middle), and period derivative (bottom) as a function of the orbital period. Colored lines show He WD models labeled by the initial central specific entropy of the donor in units of $\NA \kB$, whereas the grey line shows a He star model similar to the $0.37 \, \msun$ model in \cite{Bauer2021}. The observed properties of ZTF J0127+5258 are shown in blue lines with uncertainties bracketed by blue regions. For the He WD $4.0$ model, we show the full evolution past period minimum, with arrows in the top panel indicating the flow of time. For this model, we also show the estimated effects of tidal heating, by adding $\Etide$ (dashed line) or $\Eheat$ (dot-dashed line) to the intrinsic luminosity of the model and re-calculating the effective temperature via the Stefan-Boltzmann law. }
    \label{fig:mesa}
\end{figure*}

Higher-entropy (less degenerate) He WDs have larger radii and therefore initiate mass transfer at longer periods. As their non-degenerate outer layers become lost, their radii and thus the orbit shrink in response, and $\mdot$ reaches a peak at period minimum \citep{Kaplan2012}. Since $\mdot \propto \Jdotgr$ and longer periods have smaller $|\Jdotgr|$, higher-entropy He WDs have smaller peak $\mdot$ \citep{Deloye2007,Wong2021}. This is shown by the top panel of Figure \ref{fig:mesa} for several models with an initially $0.23 \, \msun$ donor and $0.75 \, \msun$ accretor (chosen to fit the donor orbital velocity), and different initial donor entropies ($s/(N_{\mathrm{A}} k_{\mathrm{B}}) = 3.6 - 4.0$, where $\NA$ is Avogadro's number and $\kB$ is Boltzmann constant). Except for the highest entropy model, we show only the evolution prior to period minimum. We also omit a portion of the evolution not relevant to this paper where model convergence is difficult due to an equation of state blend between Skye \citep{Jermyn2021} and CMS \citep{Chabrier2019}. 

Based on the requirement that the donor has to begin Roche-filling at $\Porb > 13.7 \, \mathrm{min}$, a low-entropy He WD can be ruled out. A He WD with $s/(\NA \kB) \approx 4.0$ and an initial mass of $0.23 \, \msun$ can furthermore yield good agreement with the measured $\dot{P}$ at $13.7 \, \mathrm{min}$, though other combinations of donor mass, accretor mass, and donor entropy may also provide a good fit. 

In addition, the high measured effective temperature of ZTF J0127+5258 favors a high-entropy He WD donor. As shown in the middle panel of Figure \ref{fig:mesa}, a higher-entropy donor has higher $\Teff$ at contact. Upon mass transfer, $\Teff$ decreases, and the $s/(\NA \kB) = 4.0$ model appears slightly too cool for ZTF J0127+5258 at $13.7 \, \mathrm{min}$, while the $s/(\NA \kB) = 3.8$ model shows a good fit. A donor initially with mass $\gtrsim 0.2 \, \msun$ is also favored, since lower mass donors never reach an effective temperature of 16400 K. However, irradiation by the accretor and tidal heating, which are ignored in our modeling, may be at work. 

Following \cite{Burdge2019b}, we estimate the effects of tidal heating on the donor surface temperature. 
Assuming that the donor is spinning synchronously with the orbit, and is rigidly rotating, the energy flux in the donor due to tides is $\Etide \approx 4 \pi^{2} I \dot{P} / P^{3}$, where $I$ is the moment of inertia of the donor. \cite{Fuller2013} found that below a critical period $\Pcrit \approx 1 \, \mathrm{hr}$, a detached He WD becomes nearly but not fully synchronized. The tidal energy flux contributes mostly to spinning up the He WD, and the rest to heat dissipation. The tidal heating rate is reduced to $\Eheat \approx \Etide (P/\Pcrit)^{19/18}$. We assume that tidal heat is dissipated near the surface of the He WD, such that its surface luminosity, $L_{\mathrm{s}}$, is obtained by adding $\Eheat$ to the intrinsic luminosity of the model. The estimated surface temperature is then given by $\Teff = \left[ L_{\mathrm{s}} / ( 4 \pi R^{2} \sigma_{\mathrm{sb}} ) \right]^{1/4}$. This is shown as a dot-dashed line in Figure \ref{fig:mesa}, for the $s/(\NA \kB) = 4.0$ model. If we choose to use $\Etide$ instead, as an upper limit to the effects of tidal heating, the surface temperature is increased to the dashed line. While tidal heating may raise the temperature of the donor to that observed, an intrinsically hot donor is likely required.

The He star model starts with $\Teff > 20{,}000\,\rm K$ before it begins donating He-dominated material. After donating enough mass to fall below the threshold for core He burning, it evolves adiabatically and its $\Teff$ decreases into the range observed for the donor in ZTF J0127+5258. However, a He star donor can only donate enough mass to reach this lower temperature after completely exhausting its residual H envelope, which always resides at mass coordinates outside $m > 0.3\,\msun$. Therefore, a model with an He star donor cannot be made consistent with the observation of residual H in the donor atmosphere at the observed $\Teff$ and $\Porb$ of ZTF J0127+5258.

The period derivative is shown in the bottom panel of Figure \ref{fig:mesa}. As discussed in Section \ref{sec:mdot_pdot}, this is influenced by $\mdot$. 
The models that are explored show $\log( \mdot / (\msunyr) )$ between $-8.0$ and $-8.5$ when they shrink to $\Porb = 13.7 \, \mathrm{min}$ from a wider orbit, in agreement with the low-$\mdot$ prior. 
However, we caution that these models give a wide range of predictions for $\mdot$ and $\dot{P}$. It is possible that ZTF J0127+5258 is only coming into contact at $13.7 \, \mathrm{min}$. Near the onset of mass transfer, $\mdot$ can vary by several orders of magnitude. For $\mdot \lesssim 10^{-9.5} \, \msunyr$, the effects of mass transfer can be negligible to $\dot{P}$. In this case, the measured donor orbital velocity and $\dot{P}$ constrain the donor mass to be $\approx 0.16 - 0.18 \, \msun$. A high-entropy $0.18 \, \msun$ He WD model in $\mesa$ can reach an effective temperature of $\approx 13000 \, \mathrm{K}$, so this scenario can match observations only if the donor surface is significantly heated by tidal dissipation or irradiation. We further note that a H-rich envelope can modify $\dot{P}$ and the mass transfer history \citep{Kaplan2012}. 

These high-entropy He WD donors have been predicted by population synthesis \citep[see Figure 1 of ][using the results of \citealt{Nelemans2001}]{Deloye2007}\footnote{For comparison, our $0.23 \msun$ He WDs with $s/(\NA \kB) = 3.8 - 4.0$ have $\log_{10}(T/\mathrm{K} )\approx 7.8$ and $\log_{10}(\rho/\mathrm{g} \, \mathrm{cm}^{-3} ) \approx 4.7-4.9$ at contact.}, and require a common envelope origin. Little time ($\approx 10^{7} \, \mathrm{yr}$) is allowed for the donor to cool to lower entropies before coming into contact again, so the post-common envelope period has to be short (30-50 min). Interestingly, extremely low-mass (ELM) WD binaries have been suggested to form at short periods \citep[lognormal distribution with mean of 30 min, albeit with high uncertainty; ][]{Brown2016}. 

\subsection{Fate of system}

ZTF J0127+5258 is currently undergoing rapid orbital decay and will either merge or reach a period minimum and begin evolving to longer orbital periods in the next one to two million years. 

Whether it merges depends in part on whether direct impact accretion can be avoided, as it increases angular momentum loss from the orbit to the accretor and tends to destabilize the system. 
Direct impact accretion likely occurs if the minimum distance that the accretion stream passes from the accretor, $\Rmin$, is smaller than the accretor radius, $\RWD$. 
For $\mesa$ He WD models (made with no H but same donor entropy as those presented in Sec \ref{sec:MESA models}) with an initially $0.75 \, \msun$ accretor, we compare $\Rmin$ \citep[][fit to \citealt{Lubow1975}]{Nelemans2001} to $\RWD$ assuming a cold WD mass-radius relation for the accretor \citep[][]{Verbunt1988}. A binary with an initially $0.23 \, \msun$ He WD donor can avoid direct impact accretion if the initial donor entropy $s/(\NA \kB) \gtrsim 3.6$, since higher-entropy donors allow for wider binary separations; with an initially $0.20 \, \msun$ He WD donor, the parameter space widens to $s/(\NA \kB) \gtrsim 3.2$. 
The He star donor models also generically avoid direct impact accretion due to their larger radii and wider orbital configurations at contact \citep{Bauer2021}.
Furthermore, we model the accretor in $\mesa$ for the $0.23 \, \msun$, $s/(\NA \kB) = 3.9 \ \& \ 4.0$ models. The accretor radius is slightly inflated compared to that of a cold WD, but direct impact accretion is still avoided (though $\Rmin$ is only $\approx 10\%$ larger than $\RWD$ at period minimum for the $s/(\NA \kB) = 3.9 $ model). 
Thus, for He WD models that provide decent matches to observations of ZTF J0127+5258, the system may narrowly avoid direct impact accretion in the future. 

Moreover, if a He nova occurs, the accretor may undergo a double-detonation as a thermonuclear supernova \citep[e.g.,][]{Shen2014}, or the system may merge \citep{Shen2015}. In the latter case, due to the compact nature of the binary, nova ejecta can easily engulf the binary; a merger may occur since dynamical friction with the nova ejecta leads to additional orbital angular momentum loss. 
For the $0.23 \, \msun$, $s/(\NA \kB) = 3.9 $ model, we find that the initially $0.75 \, \msun$ accretor will undergo a He flash after accumulating $\approx 0.06 \, \msun$ of He. However, whether a He detonation will occur is not immediately clear. 
For the $s/(\NA \kB) = 4.0 $ model, a He flash is instead avoided since the $\mdot$ is so low that accretion is unable to heat the He shell to ignition. This model will evolve to become an AM CVn binary.

\begin{figure*}
    \centering
    \includegraphics[width=\textwidth]{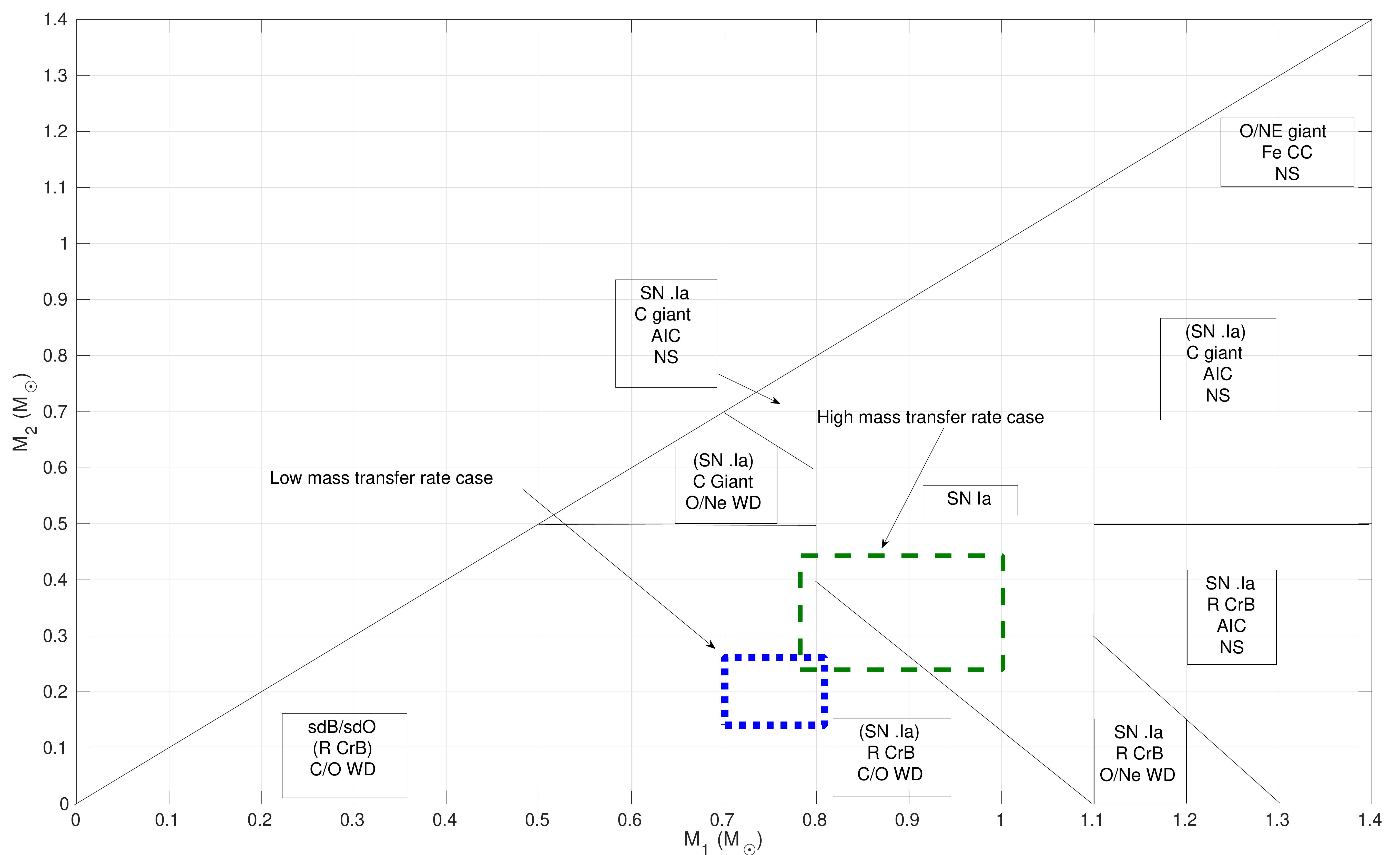}
    \caption{A diagram of the possible outcomes of double white dwarf mergers presented in \citet{Shen2015}, with the dashed blue lines indicating the mass constraints on ZTF J0127+5258 in the low mass transfer rate case, and the dashed green lines indicating the constraints in the high mass transfer rate case. If ZTF J0127+5258 fails to undergo stable mass transfer and evolve into a helium CV, this diagram indicates that it will likely form an R Coronae Borealis star which eventually cools into a white dwarf if its current mass transfer rate is on the lower end of our estimates, or will explode as a Type Ia supernova if its mass transfer rate is on the higher end of our estimates. Note that parenthetical entries in the merger outcomes may not apply to all objects in that mass range.}
    \label{fig:shen}
\end{figure*}

\section{Conclusions}

The discovery of ZTF J0127+5258 demonstrates the existence of an important new class of sources: pre-period minimum accreting binary systems with luminous donors that have orbital periods so short that they fall near the peak sensitivity of \emph{LISA}. The existence of this object suggests that at least some post-period minimum He CVs originate from the double-white-dwarf or He star channel, unless the mechanism proposed in \citet{Shen2015} prevents the system from undergoing stable mass transfer and is driven to merger instead.

As seen in Figure \ref{fig:shen}, if the system merges, it will likely form an R Crb Star \citep{Iben1984,Webbink1984,Shen2015}, which will eventually cool into a carbon-oxygen white dwarf star if its current mass transfer rate is on the lower end of our estimates. However, if the system has a high mass transfer rate, the masses we infer suggest that the system will explode in a Type Ia supernova \citep{Iben1984,Webbink1984,Shen2015}, which would make it the first gravitational wave-detectable Type Ia progenitor known. This work demonstrates the value of constraining the influence of mass transfer in the orbital evolution of these systems, as this can have major implications for the inferred past evolutionary history and predicted merger outcome. When such a source is detected by \emph{LISA}, we will still have uncertainty in the contribution of mass transfer to the orbital evolution, unless we are able to independently constrain it; however, by combining our distance estimate with the measured gravitational wave strain amplitude, we will obtain an independent constraint on the chirp mass and can infer the degree to which mass-transfer is influencing the orbital evolution.

As demonstrated in this work, these systems can exhibit rich structure in their optical lightcurves and are also sources of X-ray emission. Unlike their post-period minimum counterparts which host cold donors, binaries like ZTF J0127+5258 have donors luminous enough to directly detect in optical spectra despite the presence of a high-state accretion disk, allowing us to characterize their atmospheric composition, and measure radial velocity semi-amplitudes using phase-resolved spectroscopy. This constraint, when combined with an inclination constraint and inferred chirp mass, allows us to uniquely determine the two component masses in the system. In the \emph{LISA} era, we expect to discover many more such sources using gravitational radiation, and by then localizing them to an X-ray and/or optical/IR counterpart, we will be able to probe accretion physics in an unprecedented manner, at a large range of inclinations (as \emph{LISA} will detect both edge-on and face-on systems, and will give inclination constraints independent of the electromagnetic observations). Proposed facilities such as the \emph{Advanced X-ray Imaging Satellite} \citep{Mushotzky2019}, with its large effective area and small PSF over a moderately large FOV, will be particularly powerful tools for identifying the electromagnetic counterparts of this class of \emph{LISA} sources. This is because many of these mass-transferring sources modulate their X-flux on the orbital frequency, a frequency that will be known as a result of the \emph{LISA} detection, enabling rapid and robust association of an X-ray source within the gravitational wave error circle if there are enough X-ray photons from the source to detect this modulation.

\cleardoublepage

\section*{acknowledgments}
Kevin Burdge is a Pappalardo Postdoctoral Fellow in Physics at MIT and thanks the Pappalardo fellowship program for supporting his research. 

This work was supported in part by Chandra grant GO2-23014A.

This work was supported, in part, by the National Science Foundation through grant PHY-1748958. T.L.S.W was supported by the Gordon and Betty Moore Foundation through grant GBMF5076. T.L.S.W acknowledges use of computational facilities at UC Santa Barbara funded by NSF grant CNS 1725797, and thanks the Center for Scientific Computing for supporting this resource.

Based on observations obtained with the Samuel Oschin Telescope 48-inch and the 60- inch Telescope at the Palomar Observatory as part of the Zwicky Transient Facility project. ZTF is supported by the National Science Foundation under Grants No. AST-1440341 and AST-2034437 and a collaboration including current partners Caltech, IPAC, the Weizmann Institute of Science, the Oskar Klein Center at Stockholm University, the University of Maryland, Deutsches Elektronen-Synchrotron and Humboldt University, the TANGO Consortium of Taiwan, the University of Wisconsin at Milwaukee, Trinity College Dublin, Lawrence Livermore National Laboratories, IN2P3, University of Warwick, Ruhr University Bochum, Northwestern University and former partners the University of Washington, Los Alamos National Laboratories, and Lawrence Berkeley National Laboratories. Operations are conducted by COO, IPAC, and UW.

%

\vspace{5mm}
\facilities{PO:1.2m (ZTF), Hale (CHIMERA), Keck:I (LRIS), CXO (ACIS), Swift(XRT and UVOT)}





\appendix
\section{\emph{LISA} signal-to-noise ratio}

To estimate the \emph{LISA} SNR of the source, we follow the same procedure as presented in \citet{Burdge2020a}. In addition to this approach, we compared our SNR estimates to those estimated by the LEGWORK code \citep{Wagg2022a,Wagg2022b}. We found discrepancies between the two results, and corresponded with the authors of LEGWORK, finding several errors in both the code used to estimate SNRs in \citet{Burdge2020a}, as well as in LEGWORK, and have now converged on consistent SNR estimates after conducting this exercise. The result of these corrections reduced the estimated signal-to-noise ratios by about a factor of two in both codes (in the case of LEGWORK, the corrections applied only to the position/polarization/inclination specific sources, not the sky-averaged SNRs). The updated version of LEGWORK (0.4.0 and after) has resolved these issues.


\bibliography{sample631}{}
\bibliographystyle{aasjournal}



\end{document}